\documentclass[11pt,twoside]{article}
\usepackage[pdftex]{graphicx}
\usepackage{amsmath}
\usepackage{amssymb}
\usepackage{cite}
\usepackage{braket}

\begin{document}
\newcommand{\pst}{\hspace*{1.5em}}

\newcommand{\rigmark}{\em Journal of Russian Laser Research}
\newcommand{\lemark}{\em }

%\lhead[\fancyplain{\rigmark, {\em \lemark}}{\rigmark}]{\fancyplain{\rigmark, {\em \lemark}}{\lemark}}
%\chead{}\rhead[\fancyplain{}{\lemark}]{\fancyplain{}{\rigmark}}
%\plainfootrulewidth 0.4pt
\newcommand{\be}{\begin{equation}}
\newcommand{\ee}{\end{equation}}
\newcommand{\bm}{\boldmath}
\newcommand{\ds}{\displaystyle}
\newcommand{\bea}{\begin{eqnarray}}
\newcommand{\eea}{\end{eqnarray}}
\newcommand{\ba}{\begin{array}}
\newcommand{\ea}{\end{array}}
\newcommand{\arcsinh}{\mathop{\rm arcsinh}\nolimits}
\newcommand{\arctanh}{\mathop{\rm arctanh}\nolimits}
\newcommand{\bc}{\begin{center}}
\newcommand{\ec}{\end{center}}

\thispagestyle{plain}

\label{sh}

%\lfoot[\fancyplain{\ \\[1mm] \thepage}{\ \\[1mm]\thepage}]{\fancyplain{}{}}

\begin{center} {\Large \bf
\begin{tabular}{c}
ENTROPY DYNAMICS IN THE SYSTEM
\\[-1mm]
OF INTERACTING QUBITS
\end{tabular}
 } \end{center}

\bigskip

\bigskip

\begin{center} {\bf
N.S.\ Kirsanov$^{1*}$, A.V.\ Lebedev$^{2,1}$, M.V.\ Suslov$^1$,\\
V.M.\ Vinokur$^3$, G.\ Blatter$^2$, and G.B.\ Lesovik$^1$
}\end{center}

\medskip

\begin{center}
{\it
$^1$Moscow Institute of Physics and Technology \\
141700, Institutskii
Per. 9, Dolgoprudny, Moscow Distr., Russian Federation

\smallskip

$^2$Theoretische Physik, Wolfgang-Pauli-Strasse 27, ETH Z\"{u}rich,
CH-8093 Z\"{u}rich, Switzerland
}

\smallskip

$^3$Materials Science Division, Argonne
National Laboratory, 9700 S. Cass Ave., Argonne, IL 60439, USA

\smallskip

$^*$nikita.kirsanov@phystech.edu\\
\end{center}

\begin{abstract}\noindent
The classical Second Law of Thermodynamics demands that an isolated system
evolves with a non-diminishing entropy. This holds as well in quantum
mechanics if the evolution of the energy-isolated system can be described by a
unital quantum channel. At the same time, the entropy of a system evolving via
a non-unital channel can, in principle, decrease. Here, we analyze the
behavior of the entropy in the context of the $H$-theorem. As exemplary
phenomena, we discuss the action of a Maxwell demon (MD) operating a qubit and
the processes of heating and cooling in a two-qubit system. We further discuss
how small initial correlations between a quantum system and a reservoir affect
the increase in the entropy under the evolution of the quantum system.
\end{abstract}

\medskip


\section{Introduction}
\pst
The second law of thermodynamics restricts the processes that are allowed to
occur in nature.  The Kelvin-Planck formulation tells that it is not possible
to construct a device operating in a cycle that extracts heat from a single
reservoir and transforms it into work without other effects
\cite{Thomson:1982}. More specifically, Carnot describes what maximal
heat-to-work conversion-efficiency a heat engine can achieve that operates
with two reservoirs at different temperatures \cite{Carnot:1824}.  More
generally, in classical thermodynamics, the second law expresses an asymmetry
between the past and the future by stating that the total entropy of an
isolated system either always increases over time or remains constant in ideal
cases where the system is in a steady state or is undergoing a reversible
process. In the quantum case, however, this statement becomes trivial and the
search for a corresponding informative formulation is the subject of numerous
studies \cite{Horodecki}--\cite{Venturelli}

Within a quantum information theoretic setting, a series of exact mathematical
results has been developed that formulate the conditions for an evolution with
a non-diminishing quantum entropy as defined by von Neumann
\cite{Nielsen} --\cite{Holevo2}. This allowed for a further formulation of a
quantum analog of the classical $H$-theorem \cite{Lesovik}, i.e., the
conditions under which the entropy of an open quantum system remains
non-diminishing in the course of its quantum evolution. To that end, quantum
information theory introduces the concept of the quantum channel,
where the evolution of the system's density matrix $\hat\rho_0 \to \hat\rho_t$
is viewed as a trace-preserving completely positive map $\hat\rho_t=\Phi(\hat
\rho_0)$.  It turns out\,\cite{Holevo2} that for a unital channel $\Phi$,
i.e., a channel preserving the identity operator, $\Phi(\hat 1)=\hat 1$, the
entropy is always non-diminishing $S(\hat\rho_t) \geq S(\hat\rho_0)$.

In reference \cite{Lesovik}, it has been demonstrated that the unitality
condition is also a necessary one for a non-decreasing entropy, provided the
evolution of the system is restricted to a finite-dimensional Hilbert space.
The latter work also addressed the dynamics of an extended system comprising
both the quantum system involved and the reservoir. In general, the evolution
of such a system is described by a unitary operator
\begin{equation}
      \hat{U} = \sum_{i,j}\ket{\psi_{j}} \bra{\psi_{i}}\hat{F}_{ji},
\end{equation}
where $\{\ket{\psi_i}\}$ is an orthonormal basis in the Hilbert space, while
the family of operators $\hat{F}_{ji}$ acts in the Hilbert space of the
reservoir, with subscripts $i$ and $j$ referring to the initial and final
states of the system, respectively. Then, one can formulate an exact criterion
of unitality in terms of the operators $\hat{F}_{ji}$ for the quantum channel
$\Phi(\hat\rho) = \mbox{Tr}_\mathrm{res} \{ \hat{U} \hat\rho \otimes \hat\pi_0
\hat{U}^\dagger \}$ generated by the operator $\hat{U}$.  Here, $\hat\pi_0$ is
the initial density matrix of the reservoir and $\mbox{Tr}_\mathrm{res}$ is
the trace with respect to the reservoir states.  The sought criterion follows
from the expression for the matrix elements $\bigl[\Phi(\hat{1})
\bigr]_{jj^\prime} = \langle \psi_j| \Phi(\hat{1}) |\psi_{j^\prime}\rangle$,
\begin{equation}
\label{e1}
   \bigl[\Phi(\hat{1})\bigr]_{jj^\prime}-\bigl[\hat{1}\bigr]_{jj'} 
   = \sum_i \mbox{Tr}_\mathrm{res}\bigl\{ \hat\pi_0\, [\hat{F}^{\dagger}_{j'i},
   \hat{F}_{ji}] \bigr\}.
\end{equation}
Namely, \textit{if the right hand side of} Eq.\,(\ref{e1}) \textit{vanishes,
then the system's entropy is non-diminishing in the course of its evolution},
$\Delta S=S(\Phi(\hat \rho))-S(\hat \rho)\geq0$.

It turns out that, typically, the entropy corresponding to the evolution of an
energy-isolated system is non-decreasing. However, in some special cases,
even an energy-isolated system can evolve with a decreasing entropy. These
special cases are described by an evolution through a non-unital channel
conserving the average energy of the system and several realizations of such
channels have been analyzed in Ref.\ \cite{Lebedev}. In particular, it has
been demonstrated that such a non-unital dynamics involves a partial exchange
of the quantum system's state and that of the reservoir, giving rise to a
decreasing system's entropy if the initial state of the reservoir was more
pure (or of lower entropy) than the system's state. The decreasing of entropy
of the system without changing its average energy can be viewed as a result of
the action of a so-called quantum Maxwell demon (QMD) \cite{Lloyd}.

Here, we present a universal description of both, a classical \cite{Pekola}
and a quantum Maxwell demon, by considering the joint unitary evolution of a
system and some ancillary system that takes the role of the
\textit{reservoir}.  The chosen approach rests on the Stinespring theorem,
stating that the action of a quantum channel can be represented as the result
of the joint evolution of the given system with some auxiliary system,
where the choice of the latter is not unique. Here, we demonstrate that even a
single qubit can take up the role of a reservoir. We illustrate the action
of the previously derived $H$-theorem utilizing simple exemplary systems and
construct both unital and non-unital quantum channels modeling the thermodynamic
processes of heating and cooling.

The quantum channel formalism builds on the assumption that, initially, the
quantum system and the reservoir are not correlated. Since, in reality, the
complete absence of initial correlations between the given system and the
environment cannot be excluded, we also investigate into the influence of such
small initial correlations on the entropy dynamics and demonstrate that they
can slow down the entropy growth.

\section{Qubit realization of the Maxwell demon}
\pst
Within our approach, only two stages of the working cycle of the thermal
engine \cite{Pekola} are relevant for its description, namely, the measurement
of the qubit state and the subsequent feedback.  In particular, we are not
interested in the processes of preparing the qubit for the measurement,
neither in the interaction between the qubit and the thermal bath.  Such a
consideration then treats the feedback controller as a Maxwell demon (MD) and
thus, hereafter, we will refer to it as the \textit{demon}.

At the initial moment $t_0$, the qubit's state is described by the density matrix
\begin{equation}
\label{e2}
    \hat{\rho}_0=p_0 \ket{g} \bra{g}+p_1 \ket{e} \bra{e}\,,
\end{equation}
where $\ket{g}$ and $\ket{e}$ denote its ground and excited states. The
spacing between the energy levels of the qubit can be arbitrarily small. The
demon measures the qubit states and, based on the results, executes a
feedback operation. If the qubit is found in the ground state, the demon
leaves it unchanged.  If the qubit appears in the excited state, then the
demon drives it into the ground state by extracting work. In the course of the
process, the entropy of the qubit changes from $S_0=-p_0\ln{p_0}-p_1\ln{p_1}$
to zero, thus, the quantum channel describing the qubit's evolution is
non-unital. The action of the demon can be treated both, in the context of
either projective measurements and a feedback action depending on the outcome,
or using unitary operators for the evolution of the complete system- and
demon-qubit.

\subsection*{Semiclassical demon}
Let us consider a semiclassical demon which measures the qubit's states by the
standard procedure and demonstrate that the corresponding quantum channel is
non-unital.  Initially, at $t_0$, the qubit state is described by the density
matrix (\ref{e2}).  During the $[t_0,t_1]$-interval, the demon measures the
qubit state utilizing the corresponding projective operators $\hat{M}_g$ and
$\hat{M}_e$,
\begin{align}
    \label{e3}
    \hat{M}_g=&\ket{g}\bra{g}, & \hat{M}_e=&\ket{e}\bra{e}.
\end{align}
In the interval $[t_1,t_2]$, the qubit exercises the feedback operation that
depends on the measurement outcome. If the qubit is initially in its ground state,
the unitary operator describing the feedback has the form
\begin{equation}
    \label{e4}
    \hat{U}_g=\ket{g}\bra{g}+\ket{e}\bra{e},
\end{equation}
otherwise,
\begin{equation}
    \label{e5}
    \hat{U}_e=\ket{g}\bra{e}+\ket{e}\bra{g}.
\end{equation}
Thus the quantum channel describing the evolution of the qubit in the interval
$[t_0,t_2]$ assumes the form
\begin{equation}
\label{e6}
   \Phi(\hat{\rho}_0)= \hat{U}_g\hat{M}_g\hat{\rho}_0\hat{M}^{\dagger}_g 
   \hat{U}^{\dagger}_g+\hat{U}_e\hat{M}_e\hat{\rho}_0\hat{M}^{\dagger}_e 
   \hat{U}^{\dagger}_e.
\end{equation}
Checking for unitality, the equations (\ref{e3})--(\ref{e6}) produce the
result
\begin{equation}
\label{e7}
   \Phi(\hat{1})=2\cdot \ket{g}\bra{g},
\end{equation}
evidencing the non-unitality of the quantum channel.

\subsection*{Quantum demon}
In Ref.\ \cite{Oehri}, it has been demonstrated that the projective
measurement can be entirely described through the unitary evolution of an
extended system comprising the measured system and the measuring device. Thus, the action of the semiclassical demon discussed above can be described as a unitary process, provided the demon is a quantum system by itself. The operations described below act in the
two-dimensional subspace of the demon Hilbert space. Let this subspace be
endowed with the orthonormal basis $\ket0$ and $\ket1$ and let the $t_0$
density matrix of the complete qubit--demon system be 
\begin{equation}
\label{e8}
    \hat{R}_0= (p_0 \ket{g} \bra{g}+p_1 \ket{e} \bra{e})\otimes\ket{0} \bra{0}.
\end{equation}
During the interval $[t_0,t_1]$, the demon receives the information about the
qubit state. If the qubit is found in the ground state, the demon retains the
$\ket{0}$ state, otherwise, the demon transfers to the $\ket{1}$ state. The
evolution of the complete system at the stage of the measurement then is
described by the unitary operator
\begin{align}
\label{e9}
   \hat{U}_1=\ket{g} \bra{g} \otimes(\ket{0} \bra{0}+\ket{1} 
   \bra{1})+\ket{e} \bra{e} \otimes (\ket{0} \bra{1}+\ket{1} \bra{0}).
\end{align}
During the $[t_1,t_2]$ interval, the demon, based on the obtained information,
transfers the qubit to the ground state. The corresponding unitary evolution
operator is
\begin{align}
\label{e10}
    \hat{U}_2=(\ket{g}\bra{g}+\ket{e}\bra{e})
    \otimes\ket{0}\bra{0}
    +(\ket{g}\bra{e}+\ket{e}\bra{g})\otimes\ket{1}\bra{1}.
\end{align}
At the time $t_2$, the system state is defined by the matrix
\begin{align}
\label{e11}
    \hat{R}_f=\hat{U}_2 \hat{U}_1 \hat{R}_0 \hat{U}_1^\dagger \hat{U}_2^\dagger
    = \ket{g}\bra{g}\otimes(p_0 \ket{0} \bra{0}+p_1\ket{1} \bra{1}),
\end{align}
which evidences that the qubit is in the ground state.

Let us analyze this process in terms of the quantun $H$-theorem. The evolution
of the complete system from the moment $t_0$ till the moment $t_2$ is
described by the unitary operator
\begin{align}
\label{e12}
   \hat{U} = \hat{U}_2\,\hat{U}_1=\ket{g}\bra{g}\otimes\ket{0}\bra{0}
   +\ket{g}\bra{e}\otimes\ket{1}\bra{0}+\ket{e}\bra{g}\otimes\ket{1}\bra{1}
   +\ket{e}\bra{e}\otimes\ket{0}\bra{1}.
\end{align}
In an ideal case, the demon has to invest an infinitesimal negative work in
order to transfer the qubit from the excited to the ground state. This implies
that the qubit can be considered as an energy-isolated system which, however,
can get entangled with its environment, here, represented by the demon
qubit. The decrease in entropy is related to the non-unitality of the
corresponding quantum channel $\Phi$ generated by the operator (\ref{e12}).
To see that, let us use Eq.\ (\ref{e1}), where the operators $\hat{F}_{ji}$ can
be found from Eq.\ (\ref{e12}),
\begin{align*}
    \hat{F}_{gg}&=\ket{0}\bra{0}, & \hat{F}_{ge}&=\ket{1}\bra{0},\\
    \hat{F}_{eg}&=\ket{1}\bra{1}, & \hat{F}_{ee}&=\ket{0}\bra{1}.
\end{align*}
After some algebra, we find that 
\begin{equation}
\label{e13}
   \Phi(\hat{1})=2\cdot \ket{g}\bra{g},
\end{equation}
evidencing the non-unitality of the quantum channel. Note that Eq.\ (\ref{e13})
coincides with Eq.\ (\ref{e7}).

\section{Cooling and heating of two-level systems}
\pst

In this section, we discuss the processes of cooling and heating emerging in
various complex physical systems. Instead of presenting a full description of
the real processes, we consider a simplified model describing the interaction
of a two-level system with a reservoir operating in a two-dimensional subspace
of its Hilbert space.  We then focus on the interaction of two qubits. Let the
initial state of the two-qubit system be described by the density matrix
\begin{equation}
\label{eq14}
  \hat{R}_0=\frac{1}{2}(\ket{g_1}\bra{g_1}+\ket{e_1}\bra{e_1})\otimes\ket{g_2}\bra{g_2},
\end{equation}
where the indices 1 and 2 refer to the first and second qubit, respectively.
The cooling of the first qubit to zero and the heating of the second qubit to
an infinite temperature is described by a final density matrix assuming the
form
\begin{equation}
\label{e15}
  \hat{R}_f=\ket{g_1}\bra{g_1}\otimes\frac{1}{2}(\ket{g_2}\bra{g_2}+\ket{e_2}\bra{e_2}).
\end{equation}
The unitary evolution operator for this process is identical to the operator
given by equation (\ref{e12})
\begin{align}
\label{e16}
  \hat{U} =\ket{g_1}\bra{g_1}\otimes\ket{g_2}\bra{g_2}+\ket{g_1}\bra{e_1}
   \otimes\ket{e_2}\bra{g_2}
  +\ket{e_1}\bra{g_1}\otimes\ket{e_2}\bra{e_2}+\ket{e_1}\bra{e_1}\otimes
   \ket{g_2}\bra{e_2}.
\end{align}
Let us then discuss the evolution of each qubit separately. 

\subsection*{Evolution of the first qubit (cooling)}
The cooling process is accompanied by a decrease in entropy, thus the
quantum channel $\Phi^{(1)}$ describing the evolution of the first qubit
is non-unital.  To prove it, consider Eq.\ (\ref{e1}). The operators
$\hat{F}_{ji}$ act in the space of the states of the second qubit
and are found from Eq.\ (\ref{e16}),
\begin{align*}
    \hat{F}^{(1)}_{gg}&=\ket{g_2}\bra{g_2}, &
    \hat{F}^{(1)}_{ge}&=\ket{e_2}\bra{g_2},\\
    \hat{F}^{(1)}_{eg}&=\ket{e_2}\bra{e_2}, &
    \hat{F}^{(1)}_{ee}&=\ket{g_2}\bra{e_2}.
\end{align*}
We then find that $\Phi^{(1)}(\hat{1})=2\cdot\ket{g_1}\bra{g_1}$, i.e., the
channel $\Phi^{(1)}$ is indeed nonunital.

\subsection*{Evolution of the second qubit (heating)}
Since the entropy is increasing upon heating, the quantum channel $\Phi^{(2)}$,
describing the evolution of the second qubit can be unital. The corresponding
operators $\hat{F}_{ji}$ act in the Hilbert space of the first qubit and
assume the form
\begin{align*}
    \hat{F}^{(2)}_{gg}&=\ket{g_1}\bra{g_1}, &
    \hat{F}^{(2)}_{ge}&=\ket{e_1}\bra{e_1},\\
    \hat{F}^{(2)}_{eg}&=\ket{g_1}\bra{e_1}, &
    \hat{F}^{(2)}_{ee}&=\ket{e_1}\bra{g_1}.
\end{align*}
As a result, we find that the channel $\Phi^{(2)}(\hat1)=\hat1$ is unital.

\section{The effect of initial correlations}
\pst
So far, we have addressed processes that occur in the absence of correlations
between the initial state of the quantum system and the reservoir. Let us
consider then the change in entropy of a quantum system that is initially
correlated with the state of the reservoir and find, how the increase in
entropy relates to that in the uncorrelated situation.\,\footnote{In quantum information theory, such a situation is usually not investigated since in this case the evolution of the system cannot be described by a quantum channel.}

As before, we consider the first qubit as our system, while the second one
serves as a reservoir. The evolution of the complete system is again described
by the unitary operator (\ref{e12}), which, as we have seen, can describe
thermalization processes. One expects an increase in the system entropy if 
its initial entropy is less than the entropy of the reservoir. Let us first
consider uncorrelated subsystems and write the initial density matrix of
the complete system in the form
\begin{align}
\label{e17}
    \hat{R}_0^{(0)}=\hat{\rho}_0^{(0)}\otimes\hat{\pi}_0^{(0)},
\end{align}
where $\hat{\rho}_0^{(0)}$ describes the initial state of the system,
\begin{align}
\label{e18}
    \hat{\rho}_0^{(0)}=p_0\ket{g}\bra{g}+p_1\ket{e}\bra{e}.
\end{align}
The operator $\hat{\pi}_0^{(0)}$ describes the initial state of the
reservoir,
\begin{align}
\label{e19}
    \hat{\pi}_0^{(0)}=q_0\ket{0}\bra{0}+q_1\ket{1}\bra{1}.
\end{align}
The initial entropy of the system is
\begin{align}
\label{e20}
    S_0^{(0)}=-p_0\ln{p_0}-p_1\ln{p_1}.
\end{align}
The final state of the complete system is given by
\begin{align}
\label{e21}
    \hat{R}_f^{(0)}=&\hat{U}\hat{R}_0^{(0)}\hat{U}^\dagger\notag\\
    =&p_0 q_0\ket{g}\bra{g}\otimes\ket{0}\bra{0}+p_1 q_0 \ket{g}\bra{g}
     \otimes\ket{1}\bra{1}
    +p_0 q_1 \ket{e}\bra{e}\otimes\ket{1}\bra{1}+p_1 q_1 \ket{e}\bra{e}
    \otimes \ket{0}\bra{0}.
\end{align}
Taking the partial trace over the reservoir states, we arrive at the matrix
describing the system's final state,
\begin{align}
\label{e22}
    \hat{\rho}_f^{(0)}=q_0\ket{g}\bra{g}+q_1\ket{e}\bra{e},
\end{align}
Thus, the final entropy of the system is equal to the initial entropy of the
reservoir,
\begin{align}
\label{e23}
    S_f^{(0)}=-q_0\ln{q_0}-q_1\ln{q_1},
\end{align}
and the entropy of the system indeed increases,
\begin{equation}
\label{e24}
    \Delta S^{(0)}=S_f^{(0)}-S_0^{(0)}
    =-q_0\ln{q_0}-q_1\ln{q_1}+p_0\ln{p_0}+p_1\ln{p_1}>0.
\end{equation}

Let us now consider the situation where the initial states of the system and
the reservoir are classically correlated. The initial state of the system is
given by
\begin{equation}
\label{e25}
   \hat{R}_0=(1-\epsilon)\,\hat{R}_0^{(0)}\\
   +\epsilon\,(p_0\ket{g}\bra{g}\otimes\ket{0}\bra{0}
   +p_1\ket{e}\bra{e}\otimes\ket{1}\bra{1}),
\end{equation}
where $\epsilon$ is the small parameter; for $\epsilon=0$, the correlation
disappears and $\hat{R}_0$ turns into $\hat{R}_0^{(0)}$. One can easily see
that $\hat{R}_0$ has unit trace and the reduced density matrix of the initial
state of the system $\hat{\rho}_0$ coincides with $\hat{\rho}_0^{(0)}$. The
final state of the complete system assumes the form
\begin{equation}
\label{e26}
   \hat{R}_f=\hat{U}\hat{R}_0\hat{U}^\dagger=(1-\epsilon)\,\hat{R}_f^{(0)}
   +\epsilon\,(p_0\ket{g}\bra{g}+p_1\ket{e}\bra{e})\otimes\ket{0}\bra{0}.
\end{equation}
This yields the reduced density matrix of the system's final state as
\begin{align}
\label{e27}
    \hat{\rho}_f=\{q_0 + \epsilon\,(p_0 - q_0)\}\ket{g}\bra{g}
    +\{ q_1 + \epsilon\,(p_1 - q_1)\}\ket{e}\bra{e}.
\end{align}
Accordingly, the final entropy of the system is
\begin{align}
\label{e28}
    S_f=-\{q_0 + \epsilon\,(p_0 - q_0)\}\ln{\{q_0 + \epsilon\,(p_0 - q_0)\}}-\{q_1 + \epsilon\,(p_1 - q_1)\}\ln{\{q_1 + \epsilon\,(p_1 - q_1)\}}.
\end{align}
Now we can find the difference between the increase of entropy $\Delta S$ in the case of the small correlation between the initial states of the system and the reservoir and the quantity $\Delta S^{(0)}$ corresponding the initially uncorrelated states. Bearing in mind that
$p_0+p_1=q_0+q_1=1$, we expand in powers of small $\epsilon$ and find that
%определить S_f
%
\begin{align}
\label{e29}
   \Delta S - \Delta S^{(0)} = S_f - S_f^{(0)}
   =\epsilon\,(q_0-p_0)\ln{\frac{q_0}{1-q_0}}- \epsilon^2\,
   \frac{(p_0-q_0)^2}{2q_0 q_1} + o(\epsilon^2).
\end{align}
Hence, provided the initial entropy of the system is less than that of
the reservoir (e.g., $p_0>q_0\geq1/2$ or $p_0<q_0\leq1/2$), then $\Delta S -
\Delta S^{(0)} < 0$ if $\epsilon$ is sufficiently small. Therefore, the
presence of weak initial classical correlations results in a gain of
"correlated" entropy that is less than the increase in entropy of the
initially uncorrelated situation.  More complex cases involving an initial
entanglement of the system and the reservoir will be discussed in a
forthcoming publication.

\section{Summary}
\pst
In this work, we have demonstrated how the previously formulated quantum H-theorem can be used for analyzing the processes within the two-qubit system. Namely, examination of unitality can provide a rigorous approach to describe the work of a Maxwell demon and the thermal processes.

Moreover, we have uncovered the situation when the presence of slight initial correlations between the system and its environment can lead to a lesser entropy gain of the system in comparison to the case of the initially uncorrelated state.

\section*{Acknowledgments}
\pst
It is a pleasure to thank S.\,N. Filippov and the participants of the scientific seminar of the Laboratory of Quantum Information Theory (MIPT) for stimulating discussions. The research was supported by the Government of the Russian Federation
(Agreement № 05.Y09.21.0018), by the RFBR Grants No. 17- 02-00396A and
18-02-00642A, Foundation for the Advancement of Theoretical Physics "BASIS",
the Ministry of Education and Science of the Russian Federation
16.7162.2017/8.9 (A.V.L., in part related to the section 2), the Swiss
National Foundation via the National Centre of Competence in Research in
Quantum Science and Technology (NCCR QSIT), and the Pauli Center for
Theoretical Physics.  The work of V.M.V. was supported by the U.S. Department
of Energy, Office of Science, Materials Sciences and Engineering Division.

\end{document}